\documentclass{bmcart}
\usepackage{graphicx}
\usepackage{amsthm,amsmath}
\usepackage{booktabs}
\usepackage[utf8]{inputenc} 
\usepackage{amssymb}
\usepackage{diagbox}

\startlocaldefs
\endlocaldefs

\begin{document}

\begin{frontmatter}

\begin{fmbox}
\dochead{Research}


\title{MQENet: A Mesh Quality Evaluation Neural Network Based on Dynamic Graph Attention}


\author[
  addressref={aff1,aff2,aff3},                   
  email={zhxggg613@126.com}   
]{\fnm{Haoxuan} \snm{Zhang}}
\author[
  addressref={aff1,aff2,aff3},
  corref={aff1},              
  email={lihsh@th.btbu.edu.cn}
]{\fnm{Haisheng} \snm{Li}}
\author[
addressref={aff1,aff2,aff3},                 
email={linan@th.btbu.edu.cn}
]{\fnm{Nan} \snm{Li}}
\author[
addressref={aff1,aff2,aff3},                     
email={wangxc@btbu.edu.cn}
]{\fnm{Xiaochuan} \snm{Wang}}

\address[id=aff1]{
  \orgname{School of Computer Science and Engineering, Beijing Technology and Business University},          
  \city{Beijing 100048},                              
  \cny{China}                                    
}
\address[id=aff2]{%
  \orgname{Beijing Key Laboratory of Big Data Technology for Food Safety},
  \city{Beijing 100048},                              
\cny{China} 
}
\address[id=aff3]{%
	\orgname{National Engineering Laboratory For Agri-product Quality Traceability},
	\city{Beijing 100048},                              
	\cny{China} 
}


\end{fmbox}


\begin{abstractbox}

\begin{abstract} 
With the development of computational fluid dynamics, the requirements for the fluid simulation accuracy in industrial applications have also increased. The quality of the generated mesh directly affects the simulation accuracy. However, previous mesh quality metrics and models cannot evaluate meshes comprehensively and objectively. To this end, we propose MQENet, a structured mesh quality evaluation neural network based on dynamic graph attention. MQENet treats the mesh evaluation task as a graph classification task for classifying the quality of the input structured mesh. To make graphs generated from structured meshes more informative, MQENet introduces two novel structured mesh preprocessing algorithms. These two algorithms can also improve the conversion efficiency of structured mesh data. Experimental results on the benchmark structured mesh dataset NACA-Market show the effectiveness of MQENet in the mesh quality evaluation task.
\end{abstract}


\begin{keyword}
\kwd{Mesh quality}
\kwd{Graph attention}
\kwd{Mesh preprocessing}
\kwd{Structured mesh}
\kwd{Computational fluid dynamics}
\end{keyword}


\end{abstractbox}
%

\end{frontmatter}



\section{Introduction}
\label{sec:1}
Mesh generation is the first step in numerical calculations in Computational Fluid Dynamics (CFD)\cite{3}. In NASA's research report, titled ``CFD vision 2030 study: a path to revolutionary computational aerosciences''\cite{1}, mesh generation is listed as one of the six important research areas in the future. In modern CFD applications, mesh quality evaluation is vital in the mesh generation process\cite{2}. More importantly, the mesh quality greatly affects the accuracy of CFD simulation results\cite{4}, so researchers began to pay attention to the study of mesh quality evaluation.

In the traditional mesh quality evaluation, meshes usually utilize element-based mesh quality metrics. For 2D or surface meshes, there are some quality metrics such as area ratio, element aspect ratio, interior angle size, and smoothness. These metrics cannot comprehensively evaluate the quality of meshes, and there is no objective threshold to measure the quality of the mesh. Instead, the threshold is often determined based on the knowledge and experience of professionals.

With the development of deep learning, more and more models have been proposed and have been widely used in computer graphics~\cite{16,18}, computer vision~\cite{21,22} and natural language processing\cite{5,6}. Deep learning technology aims to improve automation and intelligence in various fields, replacing many manual operations. Some researchers have proposed using deep learning models to evaluate structured mesh quality\cite{24,23,27,8}. These methods use convolutional neural networks (CNNs), graph neural networks (GNNs) and other models to achieve good results on structured mesh quality evaluation tasks. However, due to the dynamic topology in different structured data, the features hidden in the meshes are difficult to be dynamically extracted. These deep learning-based evaluation techniques pay roughly the same attention to the majority of the topology in the structured mesh, resulting in a partial loss of accuracy. Thus, structured mesh quality evaluation based on deep learning is still a challenging task.

To make deep learning better adapt to the application of graph data, GNNs are proposed\cite{7}. Structured meshes can naturally form a graph, so it is very appropriate to adopt GNNs in structured mesh evaluation task. But existing algorithms for converting structured meshes into graphs cannot express the rich information in meshes well. In this paper, we employ dynamic graph attention to accomplish intelligent mesh quality evaluation. We propose MQENet, a graph attention neural network, for evaluating structured mesh quality without human interaction We also design two efficient structured mesh preprocessing algorithms for converting structured meshes into graphs, which can be adapted to our proposed graph neural network. The main contributions of this study include.

\begin{itemize}  
	\item{We propose MQENet, a structured mesh quality evaluation neural network based on dynamic graph attention. The network represents meshes as graphs and realizes the graph classification task with mesh quality labels by extracting features from structured meshes.}
	\item{We propose two novel structured mesh preprocessing algorithms, the point-based graph with proximity distance and the element-based graph with sparse operation. We improve the conversion efficiency compared to existing algorithms and allow the generated graph to have more features.} 
	\item{We evaluate MQENet on the mesh benchmark dataset NACA-Market. Experimental results illustrate the effectiveness of MQENet. Meanwhile, MQENet can process mesh data faster than other models.}
\end{itemize}

\section{Related work}
\label{sec:2}

\subsection{Mesh quality evaluation}
\label{sec:2.1}

Mesh quality evaluation has been widely studied in various ways. Since it is difficult to define an evaluation function that takes the entire mesh as input, mesh quality evaluation typically utilizes element-based mesh quality metrics. Several mesh quality evaluation metrics have been proposed. Li et al.~\cite{11} summarized the mesh quality metrics for commonly used 2D and 3D meshes. For 2D mesh cells, their focus is mainly on two metrics: shape and size. 2D meshes are mainly include unstructured meshes and structured meshes. The metrics for 2D unstructured meshes primarily include element length, aspect ratio and skewness. The aspect ratio of a 2D unstructured mesh~\cite{12} is measured as:
\begin{equation}
	\mbox{aspect ratio}=\dfrac{L_{\mbox{max}(L_{0},L_{1},L_{2})}}{4\sqrt{3}S}
	\label{eq:13}
\end{equation}
where $L_{\mbox{max}(L_{0},L_{1},L_{2})}$ is the length of the longest side of the unstructured mesh. $L_{0},L_{1},L_{2}$ are the lengths of the sides of the unstructured mesh and $S$ is the area of the unstructured mesh.

Skewness~\cite{13} is a mesh angle metric that indicates how close a mesh cell is to an ideal cell:
\begin{equation}
	\mbox{skewness}=\mbox{max}[\dfrac{Q_{\mbox{max}}-Q_{\mbox{ideal}}}{180-Q_{\mbox{ideal}}},\dfrac{Q_{\mbox{ideal}}-Q_{\mbox{min}}}{Q_{\mbox{ideal}}}]
		\label{eq:14}
\end{equation}	
where $Q_{\mbox{max}}$ and $Q_{\mbox{min}}$ are the maximum and minimum angles in the mesh element, respectively. $Q_{\mbox{ideal}}$ is the angle for an ideally shaped mesh element.

Although these mesh quality metrics are widely used in practice, their thresholds are often highly subjective, and it is difficult to evaluate the quality of a mesh cell from a comprehensive perspective.

In addition to the above mesh quality metrics, traditional machine learning methods are also used to evaluate mesh quality. Chetouani~\cite{14} proposed a 3D mesh quality metric based on feature fusion, which evaluates the 3D mesh quality by using the Support Vector Regression (SVR) model. Combined with the specified mesh quality metrics and geometric attributes, SVR is used to predict quality scores. Sprave et al.~\cite{15} extract low-level attributes through the neighborhood graph of the mesh. They consider the quality index of a mesh element by evaluating and summarizing the neighborhood quality index value. However, traditional machine learning methods generally have weak generalization, making it difficult to deal with data of different distributions after training.

In recent years, deep learning has been applied in various fields. To solve the above shortcomings, Chen et al.~\cite{24} proposed GridNet, a structured mesh quality evaluation based on CNNs. They also propose the structured mesh dataset NACA-Market. GridNet takes the mesh file as input and automatically evaluates the mesh quality. Chen et al.~\cite{23} also proposed an automatic hexahedral mesh evaluation framework, MVENet. It takes hexahedral mesh data as input, and studies the effect of mesh point distribution on numerical accuracy based on region segmentation and deep neural network. They employ a supervised learning process to fit the relationship between the hexahedral meshes and their quality. Wang et al.~\cite{27} proposed GMeshNet, a graph neural network to evaluate the quality of structured meshes, which converts the mesh quality evaluation task into a graph classification task. 

To sum up, traditional mesh quality metrics have significant limitations. These metrics can only evaluate a single aspect of mesh quality and rely on the expertise of professionals. Traditional machine learning methods primarily rely on manual techniques to construct mesh features when evaluating mesh quality. Deep learning-based evaluation techniques struggle to handle dynamic topology in structured meshes, resulting in difficulties achieving satisfactory accuracy.

\subsection{Graph neural network}
\label{sec:2.2}

Recent research on analyzing graphs with machine learning has gained significant attention due to the expressive power of graphs~\cite{28}. GNNs are deep learning based methods that operate on the graph domain.

Graph Convolutional Networks (GCNs) are the core part of GNNs. GCNs exploit the graph structure and aggregate node information from neighborhoods in convolutional manners~\cite{29}. Graph Attention Network (GAT) is one of the most popular GCN architecture~\cite{20}, which is considered to be a very advanced architecture in graph representation learning. Zheng et al.~\cite{30} used GAT to obtain different levels of representations to efficiently solve natural problems. Leeson et al.~\cite{31} proposed GRAVES, an algorithm selection technique in software based on GAT, which enables neural networks to make more accurate predictions by using several attention mechanisms. Cirstea et al.~\cite{32} proposed Graph-attention Recurrent neural Networks (GRN) to achieve accurate forecasting of time series.

Graph pooling is an essential component of GNNs. In order to obtain effective and reasonable graph representations, many scholars have proposed various types of graph pooling designs. Ying et al.~\cite{33} proposed a differentiable graph pooling module called DiffPool, which can generates hierarchical representations of graphs. It can be seamlessly combined with different GNNs' architecture in an end-to-end manner. Ranjan et al.~\cite{34} proposed adaptive structure aware pooling, a sparse and differentiable pooling method that addresses the limitations of previous graph pooling architectures. Ma et al.~\cite{35} proposed a path integral-based graph neural network (PAN) for graph classification and regression tasks.

Taken together, the research on GNNs has been increasing, and it has been applied in some scenarios. Since meshes can be easily represented as graph data, it is most suitable to use GNNs to process meshes. However, to the best of our knowledge, there has been very little research on GNNs for mesh quality evaluation tasks. The potential application value of GNNs need to be further explored. Therefore, this paper proposes a structured mesh quality evaluation network based on graph attention. The proposed method improves the performance on mesh datasets compared with other approaches.

\section{Methods}
\label{sec:3}

In this section, we elaborate on the technical details of two efficient mesh preprocessing algorithms and our proposed mesh quality evaluation neural network, MQENet.

\subsection{Overview}
\label{sec:3.1}

\begin{figure}[htbp]
 	\centering{}%
 	\caption{The architecture of MQENet. MQENet takes structured meshes as input and uses two mesh preprocessing schemes, the point-based graph and the element-based graph. After preprocessing, GATv2 is used to extract mesh quality features in graph nodes. SAGPool is used to filter the most important graph nodes for mesh quality. It is worth noting that graph readout operations are performed after each graph pooling. Finally, all readout graphs are concatenated and classified to obtain the mesh quality labels.}
 	
 	\includegraphics[width=0.98\linewidth]{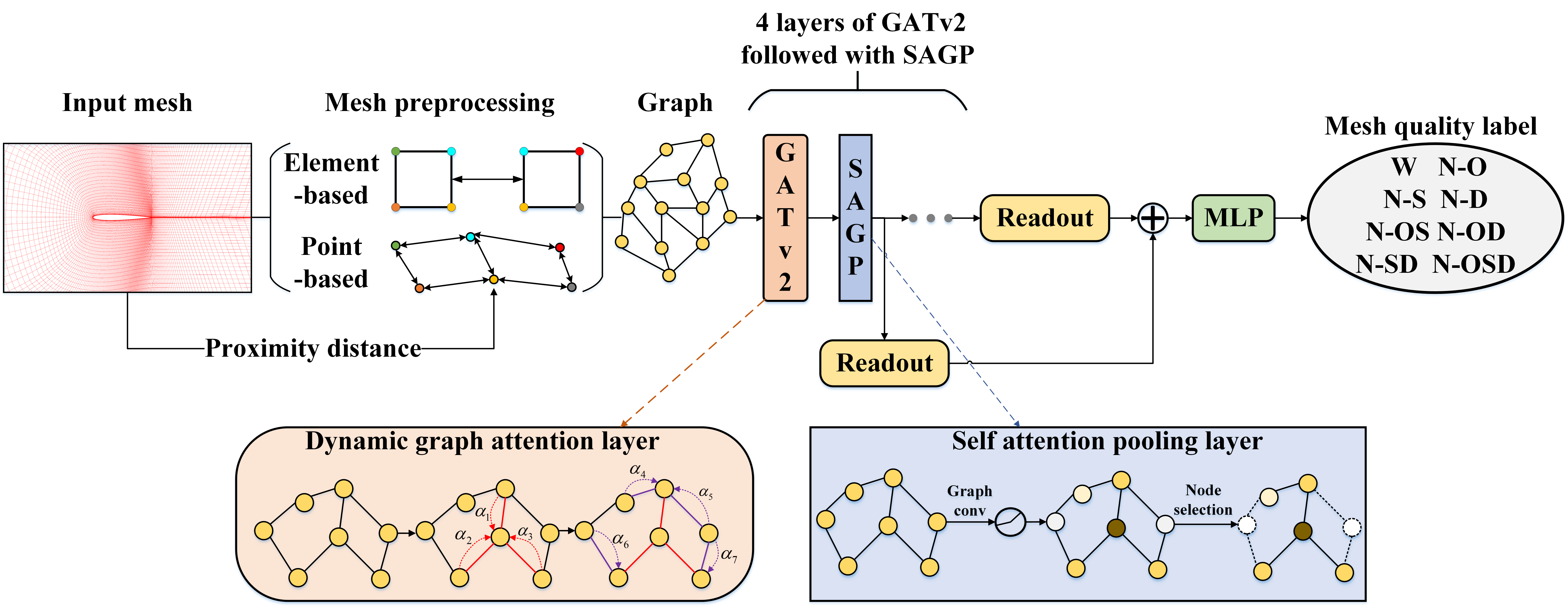}%
 	\label{fig:1}
 \end{figure}

To extract more useful high-level features from mesh elements in structured meshes, we design a neural network based on dynamic graph attention. The architecture of MQENet is shown in Figure \ref{fig:1}. 

Our proposed MQENet employs a hierarchical structure, including graph convolutional layers, graph pooling layers, graph readout operations and a multi-layer perceptron. MQENet utilizes two efficient mesh preprocessing algorithms, the point-based graph with proximity distance and the element‑based graph with sparse operation, to convert meshes into graphs. GATv2~\cite{36} and SAGPool~\cite{37} are selected as the graph convolution layer and graph pooling layer, respectively. The graph convolution layer learns the weight of neighbor nodes through a dynamic attention mechanism, which can realize the weighted aggregation of neighbor nodes. The graph pooling layer computes the projection score via a learnable vector. These scores are used to select the top-ranked nodes, thereby retaining the most valuable mesh elements. Finally, a Multi-Layer Perceptron (MLP) is used to classify the mesh quality.

\subsection{Efficient mesh preprocessing algorithm based on node and element representation}
\label{sec:3.2}
GNNs are deep learning models used for processing graph data. This paper takes structured mesh data as input, so how to convert mesh data into graph data is a challenging task. Currently, two algorithms have been proposed~\cite{27}, but they both have some problems. The point-based graph representations use only mesh nodes as graph nodes and adjacent edges as graph edges. This method does not consider the relationship between non-adjacent meshes. On the other hand, the element-based graph representation considers the diagonal elements as 1, which increases the complexity and time of mesh data conversion.

This paper proposes two efficient mesh preprocessing algorithms based on node and element representation. Inspired by Pfaff et al.~\cite{38}, the point-based graph with proximity distance for structured mesh is designed. It introduces the concept of proximity distance on the basis of the previous point-based graph. For mesh data, a mesh not only have an adjacency relationship with the directly connected mesh, but also have adjacency relationships with other meshes that have relationships but are not connected. Proximity distance edges join other mesh nodes in the graph that have spatial proximity to the current mesh node by given a fixed distance, which enhances features in mesh data. By modifying the diagonal elements, the element-based graph with sparse operation is designed, which improve the efficiency of mesh data conversion.

\subsubsection{The point-based graph with proximity distance}
\label{sec:3.2.1}

A graph is pair of $G=(V,E)$, where $V=\{v_{i}\Vert i\in N \}$ is the set of vertices from mesh nodes, N is the number of the vertices and $E = \{e_{ij}\Vert e_{ij}=(v_{i},v_{j}),(v_{i},v_{j})\in V^{2}\} $ is the set of edges from connections between meshes. For an undirected graph, $e_{ij}$ is identical to $e_{ji}$. A graph corresponding to a mesh consists of its own nodes and elements.

\begin{figure}[htbp]
	\centering{}%
	\caption{The figure shows the structured mesh of a partial wing. It can be seen that two points A,B are completely unconnected if the proximity distance is not introduced.}
	
	\includegraphics[width=0.96\linewidth]{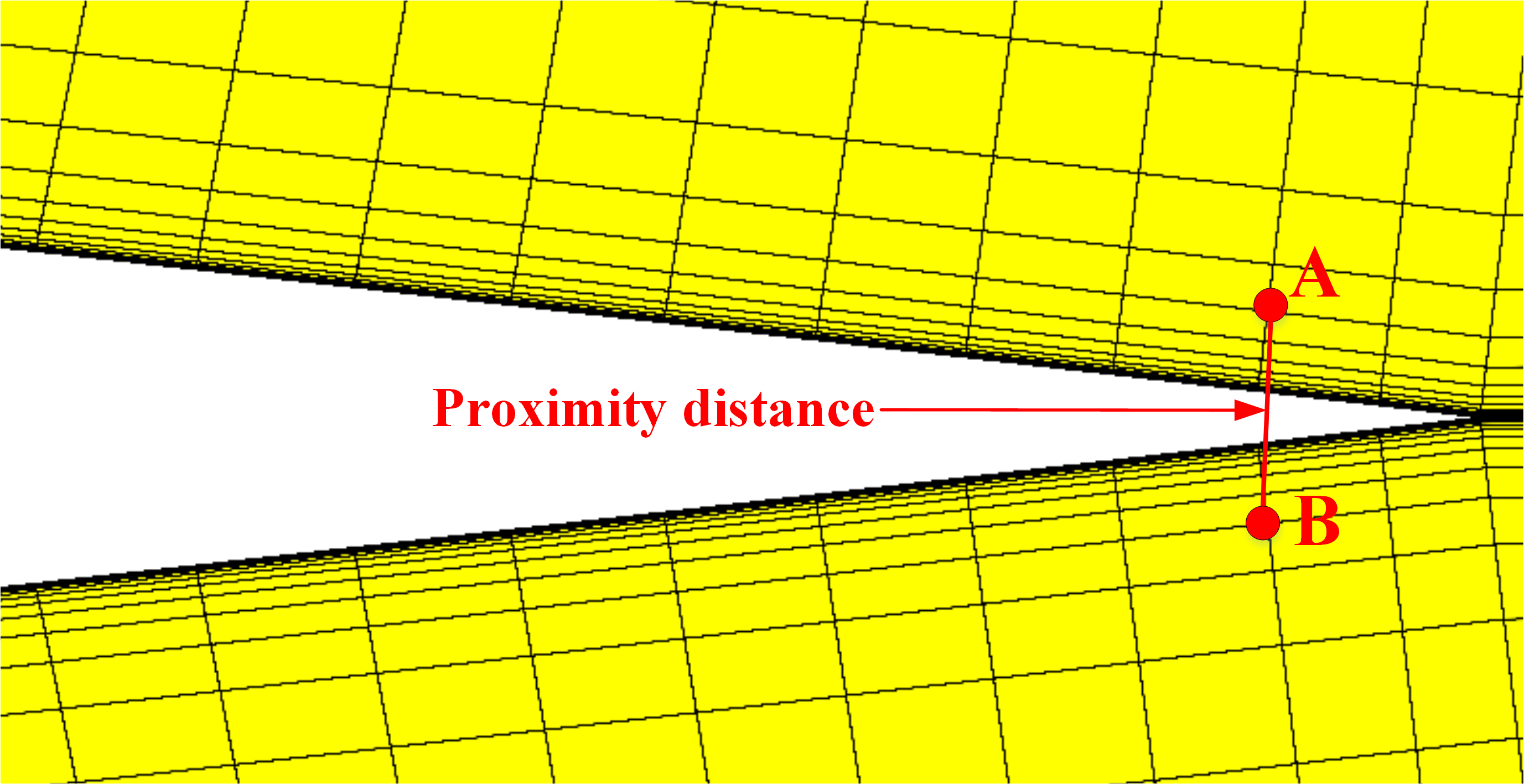}%
	\label{fig:2}
\end{figure}

We obtain the feature matrix $X_{S}\in\mathbb{R}^{4N*3} $ and the adjacency matrix $A_{S}\in\mathbb{R}^{4N*4N}$ by analyzing the coordinates in the mesh file, where $N$ is the number of meshes. Proximity distance edges are created by spatial proximity: that is, given a fixed value $r_{P}$ on the order of the smallest mesh edge lengths, we add a proximity distance edge between nodes $i$ and $j$ if $\vert dist_{ij} \vert <r_{P}$. So we can get the proximity distance matrix $A_{P}$.The input feature matrix $X$ and adjacency matrix $A$ of mesh are generated by
\begin{equation}
	X=X_{S},A=A_{S}\cup A_{P}
	\label{eq:1}
\end{equation}

This encourages using proximity distance edges to pass information between nodes that are spatially close, but distant in the mesh domain, as shown in Figure \ref{fig:2}. We use a coordinate-based sparse matrix to store the obtained feature matrix $X$ and adjacency matrix $A$, which can improve the mesh processing speed in MQENet.

\subsubsection{The element‑based graph with sparse operation}
\label{sec:3.2.2}

In addition to using mesh nodes as graph nodes, using mesh elements (mesh cells) as graph nodes is also a representation method. Wang et al.~\cite{27} proposed a graph representation technique based on mesh elements. However, they set the diagonal elements of the adjacency matrix to 1 when processing meshes, which undoubtedly increases the subsequent calculation cost. This paper improves the method of Wang et al. by modifying the diagonal elements to 0 and proposes a graph representation method based on mesh elements with sparse operations.

First, the feature matrix $X\in\mathbb{R}^{N*f}$ and the adjacency matrix of mesh nodes $A_{N}\in\mathbb{R}^{4N*4N}$ based on point are obtained from the raw mesh file, where $f$ is the number of feature (we found if the mesh is a structured mesh, $f=6$). Then, we suppose there is an element management matrix $E$, where $E=[e_{ij}]\in\mathbb{R}^{4N*N} $, and $e_{ij}=1$ if node $i$ in element $j$, otherwise $e_{ij}=0$. Finally, we can obtain the strength matrix between two mesh elements:
\begin{equation}
	S= E^{T}A_{N}E
	\label{eq:2}
\end{equation}

If two structured mesh share one edge, the strength is 6. So the adjacency matrix of mesh elements $A$ is
\begin{equation}
	A_{ij}= \left\{ \begin{array}{cc} 
	1, &\mbox{if}S_{ij}=6 \\
	0, &\mbox{otherwise}
\end{array}
	\right.
	\label{eq:3}
\end{equation}
\begin{figure}[htbp]
	\centering{}%
	\caption{The process of converting mesh elements to graph data. Each mesh cell (mesh element) is treated as a node in the graph. Adjacency relationship only occurs when two meshes share an edge.}
	
	\includegraphics[width=0.98\linewidth]{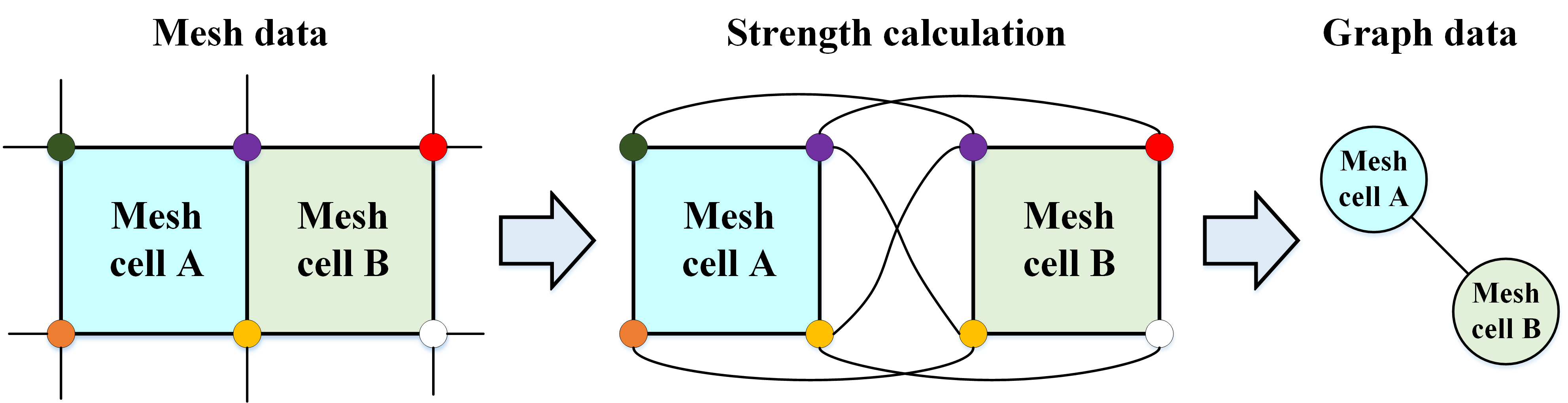}%
	\label{fig:3}
\end{figure}

As shown in Figure \ref{fig:3}, this paper takes two structured meshes as an example to describe the calculation process of the adjacency matrix of mesh elements $A$ in detail. Since the calculation does not involve mesh node coordinates, no coordinates are assigned to each node here. Among them, the adjacency matrix of mesh nodes $A_{N}$ and the element management matrix $E$ are
\begin{equation}
A_{N}=\begin{bmatrix}
	0 & 1 & 0& 1& 0& 0& 0& 0& 0\\
	1 & 0& 1& 0& 1& 0& 0& 0& 0\\
	0 & 1 & 0 & 0 & 0 & 1 & 0 & 0 & 0\\
	1 & 0 & 0 & 0 & 1 & 0 & 1 & 0 & 0\\
	0 & 1 & 0 & 1 & 0 & 1 & 0 & 1 & 0\\
	0 & 0 & 1 & 0 & 1 & 0 & 0 & 0 & 1\\
	0 & 0 & 0 & 1 & 0 & 0 & 0 & 1 & 0\\
	0 & 0 & 0 & 0 & 1 & 0 & 1 & 0 & 1\\
	0 & 0 & 0 & 0 & 0 & 1 & 0 & 1 & 0
\end{bmatrix}
,
E=\begin{bmatrix}
1 & 0 & 0 & 0\\
1 & 1 & 0 & 0\\
0 & 1 & 0 & 0\\
1 & 0 & 1 & 0\\
1 & 1 & 1 & 1\\
0 & 1 & 0 & 1\\
0 & 0 & 1 & 0\\
0 & 0 & 1 & 1\\
0 & 0 & 0 & 1
\end{bmatrix}
	\label{eq:4}
\end{equation}

Through equation (\ref{eq:2}), the strength matrix can be obtained as
\begin{equation}
	S= E^{T}A_{N}E =\begin{bmatrix}
		8 & 6 & 6 & 4\\
		6 & 8 & 4 & 6\\
		6 & 4 & 8 & 6\\
		4 & 6 & 6 & 8
	\end{bmatrix}
	\label{eq:5}
\end{equation}

Through equation (\ref{eq:3}), the adjacency matrix of mesh elements in figure \ref{fig:3} can be obtained as
\begin{equation}
	A =\begin{bmatrix}
		0 & 1 & 1 & 0\\
		1 & 0 & 0 & 1\\
		1 & 0 & 0 & 1\\
		0 & 1 & 1 & 0
	\end{bmatrix}
	\label{eq:6}
\end{equation}

The mesh preprocessing algorithm proposed in this paper can convert a data with 30,400 meshes into graph data in 0.73 seconds (the algorithm designed by Wang et al. takes 0.79 seconds).

\subsection{Dynamic graph attention layers}
\label{sec:3.3}
Different from the tasks at the node level, the graph classification task requires global information of the graph data. This information includes both the structural information of the graph and the attribute information of each node. 

In this paper, GATv2, based on the dynamic graph attention, is selected as the graph convolution layer in MQENet. The difference from GAT is that the dynamic graph attention assigns different attention scores for different nodes, while the scores obtained by the static attention mechanism are exactly the same. This shortcoming makes it difficult for GAT to distinguish the quality of different structured meshes and even hinders GAT from fitting the training data.

Some GNNs' architecture regard the weights of all neighbors as the same. To solve this problem, GAT uses the score function $\alpha:\mathbb{R}^{d}\times\mathbb{R}^{d}\rightarrow\mathbb{R}$ to score node pairs $(h_{i},h_{j})$:
\begin{equation}
	\alpha(h_{i},h_{j})=\mbox{LeakyReLU}(b^{T}\cdot[Wh_{i}\Vert Wh_{j}])
	\label{eq:7}
\end{equation}
where $b\in\mathbb{R}^{2d'},W\in\mathbb{R}^{d'\times d}$ are learnable parameters, $\Vert$ denotes vector concatenation and LeakyReLU is a type of activation function based on a ReLU.

We can obtain a more expressive dynamic attention by modifying the order of operations in the GAT. The problem with the GAT is that $W$ and $b$ in its scoring function are used for calculations in sequence, and they are likely to collapse into a linear layer. To solve this problem, we can move $b^{T}$ out of the non-linear result before performing the operation:
\begin{equation}
	\alpha(h_{i},h_{j})=b^{T}\mbox{LeakyReLU}(W\cdot[h_{i}\Vert h_{j}])
	\label{eq:8}
\end{equation}

In this way, GATv2 has powerful feature extraction capabilities for dynamic topology in structured meshes. For any two structured mesh elements or nodes, we can get the corresponding dynamic attention score $\alpha$, as shown in Figure \ref{fig:1}. Additionally, LayerNorm~\cite{39} and LeakyReLU are performed. The feature matrix $X'_{h}\in\mathbb{R}^{N\times n}$ input to the next pooling layer is:
\begin{equation}
	X'_{h}=\mbox{GATv2}(\mbox{LeakyReLU}(\mbox{LayerNorm}(X_{h})),A_{h})
	\label{eq:9}
\end{equation}
where $X_{h}\in\mathbb{R}^{N\times m}$ and $A_{h}\in\mathbb{R}^{N\times N}$ are the feature matrix and the adjacency matrix from structured meshes, respectively. $N$ is the number of the graph nodes, $m$ is the number of input features and $n$ is the number of output features.

Since GNNs with a number of network layers (more than two layers) are very prone to over-smoothing problems, we added a residual connection mechanism~\cite{43} in the graph convolutional layer to prevent this problem:
\begin{equation}
	X'_{h}=X'_{h}+X_{h}
	\label{eq:15}
\end{equation}

\subsection{Self attention pooling layers}
\label{sec:3.4}

Most graph pooling techniques need to calculate the distribution matrix of nodes in the process of clustering. However, the space-time complexity of the distribution matrix calculation algorithm is proportional to the nodes and edges of the graph. Mesh data may contain a large number of nodes and elements, which makes it difficult for most graph pooling techniques to be applied to mesh quality evaluation tasks. Thus, we adopt SAGPool to pool the graph data. 

As shown in Figure \ref{fig:1}, SAGPool adaptively learns the importance of nodes from the graph through graph convolution, and then uses the TopK~\cite{45} to discard nodes. Specifically, aggregation operations are used to assign an importance score to each node. 
\begin{equation}
Z=\mbox{tanh}(D_{h}^{-\dfrac{1}{2}}A_{h}D_{h}^{-\dfrac{1}{2}}X'_{h}W_{att})
	\label{eq:10}
\end{equation}
where tanh represents the activation function, $D_{h}\in\mathbb{R}^{N\times N}$is the degree matrix of $A_{h}$ and $W_{att}\in\mathbb{R}^{n\times 1}$ represents the weight parameter.

The pooling operation can be performed according to the importance score and the topology of the graph. Based on the score calculated by Equation \ref{eq:10}, only $\lceil kN\rceil$ nodes are reserved.
\begin{equation}
	idx=\mbox{top-rank}(Z,\lceil kN\rceil),Z_{\mbox{mask}}=Z_{\mbox{idx}}
	\label{eq:11}
\end{equation}
where $k\in(0,1]$ is the pooling ratio.

Repeatedly stacking SAGPool can perform graph pooling, and finally reduce the dimensions of each graph to the same dimension. The reconstructed feature matrix and adjacency matrix are obtained by
\begin{equation}
	X_{h+1}=X'_{h[idx,:]}\odot Z_{\mbox{mask}},A_{h+1}=A_{h[idx,idx]}
	\label{eq:12}
\end{equation}

\subsection{Graph readout operations}
\label{sec:3.5}

To read out graph data from MQENet, we make graph data more complete by concatenating the global average readout operation and the global maximum readout operation. Due to the different node positions in the graph, it is difficult to obtain an accurate representation of each node. We choose JK-net~\cite{40} to obtain an accurate representation at different levels and realize the flexible use of different neighborhood ranges for each node to achieve better structure-aware representation. The final representation is the input to an MLP for classification. Additionally, BatchNorm~\cite{41} and LeakyReLU are adopted to stabilize the training process.


\section{Experiments}
\label{sec:4}

In this section, we evaluate MQENet on the benchmark structured mesh datasets NACA-Market. The datasets, implementation details and experimental results are introduced.

\subsection{Datasets}
\label{sec:4.1}
\begin{figure}[htbp]
	\centering{}%
	\caption{Examples of meshes in NACA-Market. Red represents high quality mesh cells. Blue represents low quality mesh cells. The green part in (c) represents that the mesh smoothness is poor.}
	
	\includegraphics[width=0.97\linewidth]{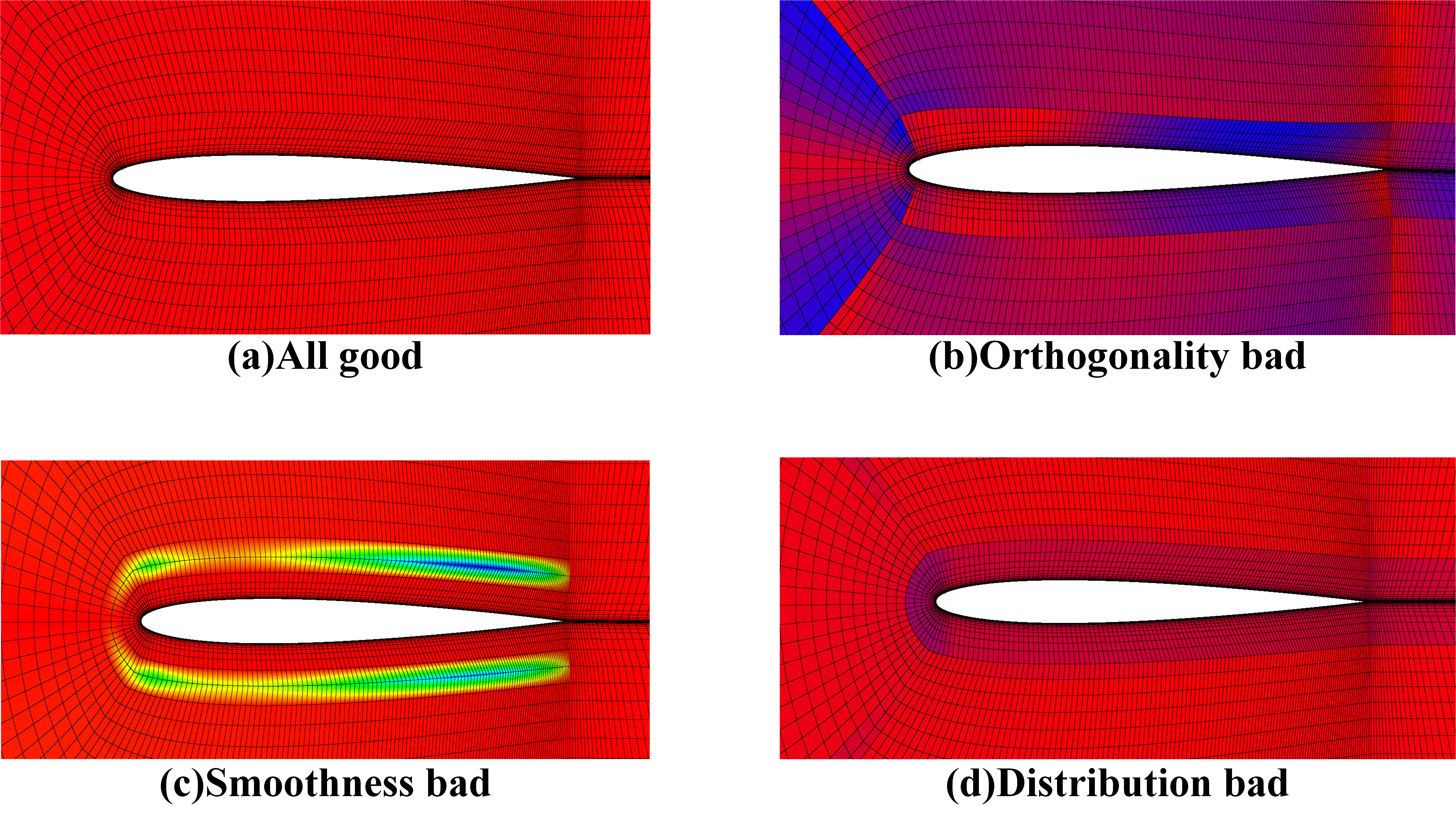}%
	\label{fig:4}
\end{figure}

Currently, there are very few benchmark structured mesh datasets available for mesh evaluation tasks. Only two datasets, NACA-Market and AirfoilSet, exist. Among them, only NACA-Market is a public dataset. So we choose the NACA-Market dataset to evaluate MQENet. NACA-Market is a mesh quality evaluation dataset of NACA0012 airfoils with eight labels, totaling 10,240 meshes. Each subset of 1,024 meshes corresponds to airfoils of one size, with a total of ten sizes. All the meshes belong to eight different quality labels, with an average of 1,280 meshes per label. The eight labels consist of three structured mesh quality evaluation metrics, which are orthogonality, smoothness and distribution, as shown in Figure \ref{fig:4}. In the experiments, we use ten NACA-Market subsets, which include 1,024 meshes and 8 label, to train MQENet separately and evaluate its performance.

\subsection{Implementation details}
\label{sec:4.2}
After converting structured meshes into graphs, we set the division ratio of the training data, validation data and test data to 60\%,20\%,20\%, and shuffle the data to ensure the distribution. We set the number of input features, output features, network layers and hidden layer units to 6, 8, 4 and 12, respectively. We select AMS-Grad~\cite{42} with an initial learning rate of 1e-2 as the optimizer, and dynamically decrease the learning rate according to the training situation. Furthermore, we use negative log likelihood loss as the loss function and add L2 regularization with 1e-4 weight decay. The batch size is set to 32. The pooling ratio for all pooling layers are set to 0.3. The experiments are carried out on an NVIDIA Tesla A100-40G.

We use the gradient clipping technique~\cite{44} to solve the gradient explosion problem that often occurs in neural networks. At the same time, the early stop method is adopted in the training process to find out the problems in the training process in time. 

\subsection{Network evaluation results}
\label{sec:4.3}
First, we evaluate the capabilities of MQENet on the test set of the NACA-Market dataset. The quality identification results of structured meshes with different characteristics are shown in Table \ref{tab:2}. Among them, W represents the mesh without defects, N-O represents the mesh with poor orthogonality, N-S represents the mesh with poor smoothness, N-D represents the mesh with poor distribution, N-OS represents the mesh with poor orthogonality and smoothness and N-OSD represents the mesh with poor orthogonality, smoothness and distribution. It can be seen that MQENet achieve very good results. The accuracy is all above 81\% on eight labels of structured meshes, and the highest can reach 83.31\%. This illustrates the effectiveness of MQENet for structured mesh datasets. 

\begin{table}[h!]
	\setlength{\tabcolsep}{1.5mm}
	\caption{Confusion matrix of the MQENet on NACA-Market. The diagonal elements represent the percentages for which the predicted label is equal to the true label for different structured meshes.}
	\begin{tabular}{ccccccccc}
		\toprule
		Labels& W (\%) & N-O(\%) & N-S (\%) & N-D (\%) &  N-OS(\%) &  N-OD(\%) &  N-SD(\%) &  N-OSD(\%)  \\ 
		\midrule
		W& \textbf{83.31} & 0.00 &2.59&12.91&0.00&0.00&1.19&0.00\\
		N-O &0.00&\textbf{82.96}&0.00&0.00&7.30&9.74&0.00&0.00 \\
		N-S &  9.52&0.00&\textbf{81.64}&0.00&7.76&0.00&1.08&0.00\\ 
		N-D&0.00&0.00&0.00&\textbf{82.48}&0.00&10.01&7.51&0.00 \\
		N-OS & 0.00&10.74&7.05&0.00&\textbf{82.21}&0.00&0.00&0.00\\
		N-OD &0.00& 9.67&0.00&0.00&0.00&\textbf{82.79}&0.00&7.54 \\ 
		N-SD&0.00&0.00& 10.42& 3.03&0.00&0.00&\textbf{81.38}&5.17  \\
		N-OSD &0.00&2.93&0.00&0.00&0.00&9.05&5.11&\textbf{82.91}\\
		\bottomrule
	\end{tabular}

\label{tab:2}
\end{table}

Then, we can also see that when the mesh has multiple defects, MQENet has the potential to identify the result as a single defect. For example, in the N-OS results, MQENet identified 11.74\% of the meshes as N-O. There are similar problems in other labels. We believe that this is a problem of labeling in the dataset. Previous studies have directly divided the NACA-Market dataset into eight categories. In doing so, the eight categories are set as completely independent labels, and the relationship between labels are not considered, which leads to the above problems.

Finally, to prove the superiority of our method, we choose to perform comparative experiments with GMeshNet. We use two metrics, recall and accuracy, to measure the precision of the neural network. As shown in Table \ref{tab:3}, we can see that the accuracy of GMeshNet trained on NACA-Market datasets is not as good as that of MQENet. In terms of orthogonality and smoothness, the recall can reach 82.96\% and 81.64\%, which is higher than GMeshNet. In terms of test accuracy, it reaches 82.67\%, which is also better than GMeshNet. More importantly, MQENet can process mesh data faster than GMeshNet.

\begin{table}[h!]
	\caption{Recall and accuracy of MQENet and GMeshNet. For another network in comparison, the results are obtained by re-running the code on NACA-Market dataset using the same training-test partition as our method.}
	\begin{tabular}{ccc}
		\toprule
		\diagbox{Mesh property}{Network}& MQENet(\%)  &  GMeshNet(\%)  \\ 
		\midrule
		Orthogonality& 82.96   & 80.73 \\
		Smoothing & 81.64  & 78.57 \\
		Distribution &   82.48   & 82.69 \\ 
		Accuracy & 82.67  & 80.93\\ 
		\bottomrule
	\end{tabular}
	\label{tab:3}
\end{table}

In summary, compared to other methods, our proposed MQENet has better accuracy and recall. This demonstrates that MQENet can perform well on structured mesh quality evaluation tasks. The network structure we designed can well adapt to the dynamic topology in different structured meshes. By filtering important nodes, hidden features in structured meshes can be more accurately captured. So that MQENet can better classify structured meshes with different quality labels in the feature space.

\subsection{Ablation studies}
\label{sec:4.4}
In this section, to illustrate the effectiveness of MQENet, this paper performs ablation studies from three aspects: mesh preprocessing algorithms, hyper-parameters and network structure.

\subsubsection{Analysis of mesh preprocessing algorithms}
\label{sec:4.4.1}
Section \ref{sec:3.2} proposes two efficient structured mesh preprocessing algorithms. We conduct experiments on two preprocessing algorithms separately. However, we only adopt the element-based graph with sparse operation to evaluate MQENet in the experiments. Compared with the work of Wang et al., we introduce proximity distance in the point-based graph, which slightly improve accuracy, but the experimental results are still unsatisfactory. We find that using the point-based graph does not even achieve 70\% accuracy in the mesh quality evaluation task. 

The point-based graph with proximity distance is a point-level representation scheme, which cannot well represent the complex topology in structured meshes. And the mesh density at the bordering is usually high due to subsequent simulation requirements. This directly leads to numerous nodes being generated at corresponding space of point-based graphs. Training GNNs on a large number of graph nodes suffers from the neighbor explosion problem, where the dependencies of nodes grow exponentially with the number of message passing layers. But these mesh points at the bordering are often an important factor to determine the quality of structured meshes.

\subsubsection{Analysis of hyper-parameters}
\label{sec:4.4.2}

In this part, we analyze the impact of different hyper-parameters on MQENet. Choosing appropriate hyper-parameters can improve the efficiency and accuracy of neural networks. We discuss two factors, activation function and pooling ratio. The experimental results are shown in Table \ref{tab:6}.

\begin{table}[h!]
	\caption{Accuracy for different hyper-parameters}
	\begin{tabular}{ccccc}
		\toprule
		\diagbox{Pooling ratio}{Activation function}& ELU(\%)  &  ReLU(\%)  &  GeLU(\%) &  LeakyReLU(\%) \\ 
		\midrule
		0.2& 80.72   & 80.87 &80.55&81.22\\
		0.3 &81.70&81.82&81.49&82.67 \\
		0.4 & 81.95&82.36&82.15&82.78   \\ 
		\bottomrule
	\end{tabular}
	\label{tab:6}
\end{table}

The pooling ratio is a parameter for each selection of nodes in the graph pooling layer. We can see that the good results in each activation function at the pooling ratio of 0.4, which means that when the pooling ratio is larger, more features will be retained. The result is relatively poor at the pooling ratio of 0.2, which shows that deleting too many nodes can reduce the size of the neural network, but the accuracy also decreases.

The other important hyper-parameter is the activation function. It allows the neural network to learn a smooth curve to segment the plane, instead of using a complex linear combination to approximate the smooth curve. We selected four activation functions of ELU, ReLU, GeLU, and LeakyReLU for experiments. In the case of the same pooling ratio, the accuracy of using LeakyReLU as the activation function is the best. Compared to other activation functions, the LeakyReLU function allows negative values to pass, which makes the dynamic range wider and allows neurons to activate more easily. For MQENet, the LeakyReLU function can better activate the ability of neurons to capture the quality features of the structured mesh. And it is more suitable for dynamic graph attention in the model.

\subsubsection{Analysis of network structure}
\label{sec:4.4.3}

In addition to the influence of mesh preprocessing algorithms and hyper-parameters on the MQENet, the model selection of the network structure is also important. As the core part of the GNN, the graph convolutional layer is the factor most likely to have an impact on the accuracy. Here we choose several classic graph convolutional layers to compare with GATv2, namely GCN~\cite{29}, GraphConv~\cite{46} and GAT~\cite{20}. Experimental results of different networks are shown in Table \ref{tab:5}. From the results, we can see that GATv2 performs better than other graph convolutional layers on NACA-Market datasets. 

\begin{table}[h!]
	\caption{Accuracy for different networks}
	\begin{tabular}{cc}
		\toprule
		Method & Accuracy(\%) \\ 
		\midrule
		GCN & 77.87 \\
		GraphConv & 78.14 \\
		GAT & 80.93 \\
		GATv2 & \textbf{82.67} \\ 
		\bottomrule
	\end{tabular}
	\label{tab:5}
\end{table}

GATv2 can enhance the representation of hidden features in structured meshes by computing the dynamic attention score of each node. Then features related to mesh quality are mined to enable the ability to classify structured meshes with different labels. Other graph convolutional layers are difficult to deal with dynamic topology in structured meshes due to no attention mechanism or only static attention mechanism.


\section{Conclusions}
\label{sec:5}

The quality of the mesh is a critical factor in the accuracy of CFD simulations. However, traditional mesh quality metrics and learning-based mesh quality evaluation techniques have been unable to meet the industry's increasing demand for higher CFD simulation accuracy. Nodes and cells in structured meshes can naturally form a graph, so it is very appropriate to adopt graph neural networks in mesh evaluation task. Therefore, a structured mesh quality evaluation nerual network based on dynamic graph attention, MQENet, is proposed. We also design two novel structured mesh preprocessing algorithms, the point-based graph with proximity distance and the element-based graph with sparse operation, to convert meshes to graphs, which further enhance the conversion efficiency. We evaluate MQENet on the structured mesh datasets NACA-Market and demonstrate that MQENet outperforms other methods in the mesh quality evaluation task.

In future work, we will further explore neural networks suitable for point-based graphs. To address the problem of labels not being independent in NACA-Market datasets, we plan to change the eight-category problem to a multi-label problem (three labels), which may avoids the problem of mutual influence of defects.



\begin{backmatter}

\section*{Availability of data and materials}
The datasets during the current study are available in https://github.com/chenxinhai1234/NACA-Market.

\section*{Ethics approval and consent to participate}
Not applicable.

\section*{Competing interests}
The authors declare that they have no competing interests.

\section*{Consent for publication}
Not applicable.

\section*{Authors' contributions}
Haoxuan Zhang: Conceptualization, Methodology, Writing. Haisheng Li:  Funding acquisition, Project administration, Data curation. Nan Li: Formal analysis, Writing, Supervision. Xiaochuan Wang: Investigation, Writing, Supervision.

\section*{Funding}
This work is supported by the National Natural Science Foundation of China (No. 62277001, No. 62272014, No. 62201017), and Scientific Research Program of Beijing Municipal Education Commission KZ202110011017.

\section*{Acknowledgements}
This work was carried out at National Supercomputer Center in Tianjin, and the calculations were performed on Tianhe new generation Supercomputer. We also thank the Beijing Technology and Business University 2023 Postgraduate Research Ability Improvement Program Project for funding.

\section*{Authors' information}

Haoxuan Zhang, male, received his bachelor’s degree from Beijing Technology and Business University, China in 2022. He is currently pursuing a Master of Engineering degree from Beijing Technology and Business University, China. His research interests include deep learning and mesh generation.

Haisheng Li (corresponding author), male, received his Ph.D. degree in computer graphics from Beihang University, China in 2002. He is a professor in school of Computer Science and Engineering, Beijing Technology and Business University, China. His current research interests include mesh generation, computer graphics and intelligent information processing etc.

Nan Li, male, received his bachelor's and Ph.D. degrees from Beijing Jiaotong University in 2004 and 2010, respectively. He is a professor of Beijing Technology and Business University. His main research interests are design methodology, computer graphics and intelligent engineering.

Xiaochuan Wang, male, received his master's and Ph.D. degrees from Beihang University in 2012 and 2019, respectively. He is an associate professor of Beijing Technology and Business University. His main research interests are virtual reality and computer graphics.

\end{backmatter}


\begin{thebibliography}{6}
\ifx \bisbn   \undefined \def \bisbn  #1{ISBN #1}\fi
\ifx \binits  \undefined \def \binits#1{#1}\fi
\ifx \bauthor  \undefined \def \bauthor#1{#1}\fi
\ifx \batitle  \undefined \def \batitle#1{#1}\fi
\ifx \bjtitle  \undefined \def \bjtitle#1{#1}\fi
\ifx \bvolume  \undefined \def \bvolume#1{\textbf{#1}}\fi
\ifx \byear  \undefined \def \byear#1{#1}\fi
\ifx \bissue  \undefined \def \bissue#1{#1}\fi
\ifx \bfpage  \undefined \def \bfpage#1{#1}\fi
\ifx \blpage  \undefined \def \blpage #1{#1}\fi
\ifx \burl  \undefined \def \burl#1{\textsf{#1}}\fi
\ifx \doiurl  \undefined \def \doiurl#1{\textsf{#1}}\fi
\ifx \betal  \undefined \def \betal{\textit{et al.}}\fi
\ifx \binstitute  \undefined \def \binstitute#1{#1}\fi
\ifx \binstitutionaled  \undefined \def \binstitutionaled#1{#1}\fi
\ifx \bctitle  \undefined \def \bctitle#1{#1}\fi
\ifx \beditor  \undefined \def \beditor#1{#1}\fi
\ifx \bpublisher  \undefined \def \bpublisher#1{#1}\fi
\ifx \bbtitle  \undefined \def \bbtitle#1{#1}\fi
\ifx \bedition  \undefined \def \bedition#1{#1}\fi
\ifx \bseriesno  \undefined \def \bseriesno#1{#1}\fi
\ifx \blocation  \undefined \def \blocation#1{#1}\fi
\ifx \bsertitle  \undefined \def \bsertitle#1{#1}\fi
\ifx \bsnm \undefined \def \bsnm#1{#1}\fi
\ifx \bsuffix \undefined \def \bsuffix#1{#1}\fi
\ifx \bparticle \undefined \def \bparticle#1{#1}\fi
\ifx \barticle \undefined \def \barticle#1{#1}\fi
\ifx \bconfdate \undefined \def \bconfdate #1{#1}\fi
\ifx \botherref \undefined \def \botherref #1{#1}\fi
\ifx \url \undefined \def \url#1{\textsf{#1}}\fi
\ifx \bchapter \undefined \def \bchapter#1{#1}\fi
\ifx \bbook \undefined \def \bbook#1{#1}\fi
\ifx \bcomment \undefined \def \bcomment#1{#1}\fi
\ifx \oauthor \undefined \def \oauthor#1{#1}\fi
\ifx \citeauthoryear \undefined \def \citeauthoryear#1{#1}\fi
\ifx \endbibitem  \undefined \def \endbibitem {}\fi
\ifx \bconflocation  \undefined \def \bconflocation#1{#1}\fi
\ifx \arxivurl  \undefined \def \arxivurl#1{\textsf{#1}}\fi
\csname PreBibitemsHook\endcsname

\bibitem{koon}
\begin{barticle}
\bauthor{\bsnm{Koonin}, \binits{E.V.}},
\bauthor{\bsnm{Altschul}, \binits{S.F.}},
\bauthor{\bsnm{Bork}, \binits{P.}}:
\batitle{Brca1 protein products: functional motifs}.
\bjtitle{Nat. Genet.}
\bvolume{13},
\bfpage{266}--\blpage{267}
(\byear{1996})
\end{barticle}
\endbibitem

\bibitem{xjon}
\begin{bchapter}
\bauthor{\bsnm{Jones}, \binits{X.}}:
\bctitle{Zeolites and synthetic mechanisms}.
In: \beditor{\bsnm{Smith}, \binits{Y.}} (ed.)
\bbtitle{Proceedings of the First National Conference on Porous Sieves: 27-30
  June 1996; Baltimore},
pp. \bfpage{16}--\blpage{27}
(\byear{1996})
\end{bchapter}
\endbibitem

\bibitem{marg}
\begin{bbook}
\bauthor{\bsnm{Margulis}, \binits{L.}}:
\bbtitle{Origin of Eukaryotic Cells}.
\bpublisher{Yale University Press},
\blocation{New Haven}
(\byear{1970})
\end{bbook}
\endbibitem

\bibitem{schn}
\begin{bchapter}
\bauthor{\bsnm{Schnepf}, \binits{E.}}:
\bctitle{From prey via endosymbiont to plastids: comparative studies in
  dinoflagellates}.
In: \beditor{\bsnm{Lewin}, \binits{R.A.}} (ed.)
\bbtitle{Origins of Plastids},
\bedition{2nd} edn.,
pp. \bfpage{53}--\blpage{76}.
\bpublisher{Chapman and Hall},
\blocation{New York}
(\byear{1993})
\end{bchapter}
\endbibitem

\bibitem{koha}
\begin{botherref}
\oauthor{\bsnm{Kohavi}, \binits{R.}}:
Wrappers for performance enhancement and obvious decision graphs.
PhD thesis,
Stanford University, Computer Science Department
(1995)
\end{botherref}
\endbibitem

\bibitem{issnic}
\begin{botherref}
\oauthor{\bsnm{{ISSN International Centre}}}:
The ISSN register
(2006).
\url{http://www.issn.org}
Accessed Accessed 20 Feb 2007
\end{botherref}
\endbibitem

\end{thebibliography}

\newcommand{\BMCxmlcomment}[1]{}

\BMCxmlcomment{

<refgrp>

<bibl id="B1">
  <title><p>BRCA1 protein products: functional motifs</p></title>
  <aug>
    <au><snm>Koonin</snm><fnm>E V</fnm></au>
    <au><snm>Altschul</snm><fnm>S F</fnm></au>
    <au><snm>Bork</snm><fnm>P</fnm></au>
  </aug>
  <source>Nat. Genet.</source>
  <pubdate>1996</pubdate>
  <volume>13</volume>
  <fpage>266</fpage>
  <lpage>267</lpage>
</bibl>

<bibl id="B2">
  <title><p>Zeolites and synthetic mechanisms</p></title>
  <aug>
    <au><snm>Jones</snm><fnm>X</fnm></au>
  </aug>
  <source>Proceedings of the First National Conference on Porous Sieves: 27-30
  June 1996; Baltimore</source>
  <editor>Y Smith</editor>
  <pubdate>1996</pubdate>
  <fpage>16</fpage>
  <lpage>27</lpage>
</bibl>

<bibl id="B3">
  <title><p>Origin of Eukaryotic Cells</p></title>
  <aug>
    <au><snm>Margulis</snm><fnm>L</fnm></au>
  </aug>
  <publisher>New Haven: Yale University Press</publisher>
  <pubdate>1970</pubdate>
</bibl>

<bibl id="B4">
  <title><p>From prey via endosymbiont to plastids: comparative studies in
  dinoflagellates</p></title>
  <aug>
    <au><snm>Schnepf</snm><fnm>E</fnm></au>
  </aug>
  <source>Origins of Plastids</source>
  <publisher>New York: Chapman and Hall</publisher>
  <editor>R A Lewin</editor>
  <edition>2</edition>
  <pubdate>1993</pubdate>
  <fpage>53</fpage>
  <lpage>76</lpage>
</bibl>

<bibl id="B5">
  <title><p>Wrappers for performance enhancement and obvious decision
  graphs</p></title>
  <aug>
    <au><snm>Kohavi</snm><fnm>R</fnm></au>
  </aug>
  <source>PhD thesis</source>
  <publisher>Stanford University, Computer Science Department</publisher>
  <pubdate>1995</pubdate>
</bibl>

<bibl id="B6">
  <title><p>The ISSN register</p></title>
  <aug>
    <au><cnm>{ISSN International Centre}</cnm></au>
  </aug>
  <pubdate>2006</pubdate>
  <url>http://www.issn.org</url>
</bibl>

</refgrp>
} 


\begin{thebibliography}{99.}

\bibitem{3}
Lei N, Li Z, Xu Z, et al. What's the Situation With Intelligent Mesh Generation: A Survey and Perspectives[J]. IEEE Transactions on Visualization and Computer Graphics, 2023.

\bibitem{1}
Slotnick J P, Khodadoust A, Alonso J, et al. CFD vision 2030 study: a path to revolutionary computational aerosciences[R]. 2014.

\bibitem{2}
Thornburg H. Overview of the PETTT Workshop on Mesh Quality/Resolution, Practice, Current Research, and Future Directions[C]//50th AIAA Aerospace Sciences Meeting including the New Horizons Forum and Aerospace Exposition. 2012: 606.
 
\bibitem{4}
Katz A, Sankaran V. Mesh quality effects on the accuracy of CFD solutions on unstructured meshes[J]. Journal of Computational Physics, 2011, 230(20): 7670-7686.

\bibitem{16}
Li H, Zhao T, Li N, et al. Feature matching of multi-view 3d models based on hash binary encoding[J]. Neural Network World, 2017, 27(1): 95.

\bibitem{18}
Li H, Zheng Y, Cao J, et al. Multi-view-based siamese convolutional neural network for 3D object retrieval[J]. Computers and Electrical Engineering, 2019, 78: 11-21.

\bibitem{21}
Zheng Y, Zeng G, Li H, et al. Colorful 3D reconstruction at high resolution using multi-view representation[J]. Journal of Visual Communication and Image Representation, 2022: 103486.

\bibitem{22}
Li H, Zheng Y, Wu X, et al. 3D model generation and reconstruction using conditional generative adversarial network[J]. International Journal of Computational Intelligence Systems, 2019, 12(2): 697.


\bibitem{5}
Floridi L, Chiriatti M. GPT-3: Its nature, scope, limits, and consequences[J]. Minds and Machines, 2020, 30: 681-694.

\bibitem{6}
Zhang C, Zhang C, Zheng S, et al. A complete survey on generative ai (aigc): Is chatgpt from gpt-4 to gpt-5 all you need?[J]. arXiv preprint arXiv:2303.11717, 2023.

\bibitem{24}
Chen X, Liu J, Pang Y, et al. Developing a new mesh quality evaluation method based on convolutional neural network[J]. Engineering Applications of Computational Fluid Mechanics, 2020, 14(1): 391-400.

\bibitem{23}
Chen X, Liu J, Gong C, et al. MVE-Net: An automatic 3-D structured mesh validity evaluation framework using deep neural networks[J]. Computer-Aided Design, 2021, 141: 103104.

\bibitem{27}
Wang Z, Chen X, Li T, et al. Evaluating mesh quality with graph neural networks[J]. Engineering with Computers, 2022, 38(5): 4663-4673.

\bibitem{8}
Liu Z, Liu H, Chen Y, et al. Evaluating Airfoil Mesh Quality with Transformer[J]. Aerospace, 2023, 10(2): 110.

\bibitem{7}
Wu L, Cui P, Pei J, et al. Graph neural networks: foundation, frontiers and applications[C]//Proceedings of the 28th ACM SIGKDD Conference on Knowledge Discovery and Data Mining. 2022: 4840-4841.

\bibitem{11}
Li H. Finite Element Mesh Generation and Decision Criteria of Mesh Quality[J]. China Mechanical Engineering, 2012, 23(3): 368.

\bibitem{12}
Knupp P M, Ernst C D, Thompson D C, et al. The verdict geometric quality library[R]. Sandia National Laboratories (SNL), Albuquerque, NM, and Livermore, CA (United States), 2006.

\bibitem{13}
Robinson J. CRE method of element testing and the Jacobian shape parameters[J]. Engineering Computations, 1987, 4(2): 113-118.

\bibitem{14}
Chetouani A. A 3D mesh quality metric based on features fusion[J]. Electronic Imaging, 2017, 29: 4-8.

\bibitem{15}
Sprave J, Drescher C. Evaluating the quality of finite element meshes with machine learning[J]. arXiv preprint arXiv:2107.10507, 2021.

\bibitem{28}
Wu Z, Pan S, Chen F, et al. A comprehensive survey on graph neural networks[J]. IEEE Transactions on Neural Networks and Learning Systems, 2020, 32(1): 4-24.

\bibitem{29}
Kipf T N, Welling M. Semi-supervised classification with graph convolutional networks[J]. arXiv preprint arXiv:1609.02907, 2016.

\bibitem{20}
Veličković P, Cucurull G, Casanova A, et al. Graph attention networks[J]. arXiv preprint arXiv:1710.10903, 2017.

\bibitem{30}
Zheng B, Wen H, Liang Y, et al. Document modeling with graph attention networks for multi-grained machine reading comprehension[J]. arXiv preprint arXiv:2005.05806, 2020.

\bibitem{31}
Leeson W, Dwyer M B. Algorithm Selection for Software Verification using Graph Attention Networks[J]. arXiv preprint arXiv:2201.11711, 2022.

\bibitem{32}
Cirstea R G, Guo C, Yang B. Graph Attention Recurrent Neural Networks for Correlated Time Series Forecasting--Full version[J]. arXiv preprint arXiv:2103.10760, 2021.

\bibitem{33}
Ying Z, You J, Morris C, et al. Hierarchical graph representation learning with differentiable pooling[J]. Advances in Neural Information Processing Systems, 2018, 31.

\bibitem{34}
Ranjan E, Sanyal S, Talukdar P. Asap: Adaptive structure aware pooling for learning hierarchical graph representations[C]//Proceedings of the AAAI Conference on Artificial Intelligence. 2020, 34(04): 5470-5477.

\bibitem{35}
Ma Z, Xuan J, Wang Y G, et al. Path integral based convolution and pooling for graph neural networks[J]. Advances in Neural Information Processing Systems, 2020, 33: 16421-16433.

\bibitem{36}
Brody S, Alon U, Yahav E. How attentive are graph attention networks?[J]. arXiv preprint arXiv:2105.14491, 2021.

\bibitem{37}
Lee J, Lee I, Kang J. Self-attention graph pooling[C]//International Conference on Machine Learning. PMLR, 2019: 3734-3743.

\bibitem{38}
Pfaff T, Fortunato M, Sanchez-Gonzalez A, et al. Learning mesh-based simulation with graph networks[J]. arXiv preprint arXiv:2010.03409, 2020.

\bibitem{39}
Ba J L, Kiros J R, Hinton G E. Layer normalization[J]. arXiv preprint arXiv:1607.06450, 2016.

\bibitem{43}
Li G, Muller M, Thabet A, et al. Deepgcns: Can gcns go as deep as cnns?[C]//Proceedings of the IEEE/CVF International Conference on Computer Vision. 2019: 9267-9276.

\bibitem{45}
Knyazev B, Taylor G W, Amer M. Understanding attention and generalization in graph neural networks[J]. Advances in Neural Information Processing Systems, 2019, 32.

\bibitem{40}
Xu K, Li C, Tian Y, et al. Representation learning on graphs with jumping knowledge networks[C]//International Conference on Machine Learning. PMLR, 2018: 5453-5462.

\bibitem{41}
Ioffe S, Szegedy C. Batch normalization: Accelerating deep network training by reducing internal covariate shift[C]//International Conference on Machine Learning. PMLR, 2015: 448-456.

\bibitem{42}
Reddi S J, Kale S, Kumar S. On the convergence of adam and beyond[J]. arXiv preprint arXiv:1904.09237, 2019.

\bibitem{44}
Pascanu R, Mikolov T, Bengio Y. On the difficulty of training recurrent neural networks[C]//International Conference on Machine Learning. PMLR, 2013: 1310-1318.

\bibitem{46}
Morris C, Ritzert M, Fey M, et al. Weisfeiler and leman go neural: Higher-order graph neural networks[C]//Proceedings of the AAAI Conference on Artificial Intelligence. 2019, 33(01): 4602-4609.


\end{thebibliography}
\end{document}